\documentclass[aps,prx,twocolumn,superscriptaddress]{revtex4-2}

\usepackage{graphicx}% Include figure files
\usepackage{dcolumn}% Align table columns on decimal point
\usepackage{bm}% bold math
\usepackage[dvipsnames]{xcolor}
\usepackage{amsmath}

\definecolor{OliveGreen}{rgb}{0.5,0.5,0.0}
\definecolor{BrickRed}{rgb}{0.8, 0.25, 0.33}
\definecolor{Cerulean}{rgb}{0.0, 0.48, 0.65}
\definecolor{Fuchsia}{rgb}{1.0, 0.0, 1.0}
\definecolor{NavyBlue}{rgb}{0.0, 0.0, 0.5}
\definecolor{green}{rgb}{0, 1, 0}
\definecolor{orange}{rgb}{1 0.5 0}

\newcommand{\That}{\widehat{T}}
\newcommand{\Rmrca}{R_\text{MRCA}}
\newcommand{\Tmrca}{T_\text{MRCA}}
\newcommand{\figref}[1]{Fig.~\ref{#1}}
\newcommand{\eqnref}[1]{Eq.~\eqref{#1}}
\begin{document}

\title{Branching structure of genealogies in spatially growing populations and its implications for population genetics inference}% Force line breaks with \\

\author{Armin Eghdami}
  \email{armin.eghdami@berkeley.edu}
 \affiliation{Department of Physics, University of Cambridge.}%Lines break automatically or can be forced with \\

\author{Jayson Paulose}
 \email{jpaulose@uoregon.edu}
\affiliation{%
Department of Physics and Institute for Fundamental Science, University of Oregon.
}%

\author{Diana Fusco}
 \email{df390@cam.ac.uk}
\affiliation{
 Department of Physics, University of Cambridge.
}%


\begin{abstract}
Spatial models where growth is limited to the population edge have been instrumental to understanding the population dynamics and the clone size distribution in growing cellular populations, such as microbial colonies and avascular tumours. A complete characterization of the coalescence process generated by spatial growth is still lacking, limiting our ability to apply classic population genetics inference to spatially growing populations.
Here, we start filling this gap by investigating the statistical properties of the cell lineages generated by the two dimensional Eden model, leveraging their physical analogy with directed polymers. Our analysis provides quantitative estimates for population measurements that can easily be assessed via sequencing, such as the average number of segregating sites and the clone size distribution of a subsample of the population. Our results not only reveal remarkable features of the genealogies generated during growth, but also highlight new properties that can be misinterpreted as signs of selection if non-spatial models are inappropriately applied.
\end{abstract}

\maketitle

%\tableofcontents

\graphicspath{{pictures/}}

%-----------------------------------------------------------------------------------------%
%-----------------------------------------------------------------------------------------%
\section{\label{sec:section1}Introduction}
Spatial range expansions \cite{old_range_expansion_paper,1_main} are ubiquitous in nature, from microbial biofilms \cite{1_1,fusco}, developing tissues \cite{1_2} avascular tumors~\cite{1_3,1_4,1_5,1_6} to invading species and infectious diseases~\cite{Thompson2021,Real2007}. Many of these scenarios share the feature of being resources-limited \cite{1_7,1_8,2_5,1_10,1_11,1_12,1_13,1_14}, so that population growth occurs mainly as invasion of surrounding virgin territory \cite{hallatschek_eden,1_20,1_21,1_22,1_23,1_24,1_25}. When dispersal is local, these range expansions lead to a phenomenon called \textit{gene surfing}, whereby pioneering individuals at the edge of the expansion have a higher chance to contribute to the next generation~\cite{Hallatschek2008,Excoffier2008}. As a result, 
an individual's location can become a more important factor to reproductive success than its growth rate \cite{1_15,1_16,1_17,1_18}. 

It has recently been shown that gene surfing leaves a characteristic signature in the mutational spectrum of the population, identified by an excess of high frequency mutations compared to the well-mixed expectation~\cite{fusco}. This observation becomes crucial when analyzing population sequencing results, as the same signature can be mistakenly interpreted as being a result of positive selection and lead to a mis-identification of driver mutations (e.g., in cancer or drug resistance).
%How to discern then the two scenarios?
Modeling the effects of spatial structure on genealogies in growing populations, and consequently on the diagnostic outputs of genome sequencing, could point to protocols that discern the two scenarios.

  In prior efforts to link spatial growth with population structure, connections to non-equilibrium statistical mechanics have proven fruitful. Spatial population growth is a fundamentally out-of-equilibrium process driven by stochastic division and migration events at small scales that leave collective signatures at the scale of the population as a whole. Models of spatial populations harbor nonequilibrium statistical phenomena such as fixation into absorbing states~\cite{Korolev2010}, dynamic phase transitions and critical phenomena~\cite{Kuhr2011,Lavrentovich2013}, and manifestations of directed percolation~\cite{Horowitz2019}. As a result, statistical properties of the population patterns generated by the range expansion can be quantitatively linked to robust universal features of the corresponding nonequilibrium growth models~\cite{Hinrichsen2000}. A prime example is the connection between range expansions and the Kardar-Parisi-Zhang (KPZ) model of interface growth~\cite{kpz}, which established scaling exponents for the shapes of clonal domains~\cite{Derrida1991,hallatschek_proof,hallatschek_eden} and lineages~\cite{Manna1996,Cieplak1996,rand_walker} in expanding microbial populations. Such scaling rules provide a theoretical basis for predicting or interpreting population genetics quantities that can be measured using genome sequencing studies.

In the following, we systematically analyse the statistical properties of the genealogical tree generated by the Eden model, a lattice model that has successfully been used to investigate microbial colonies and tumour growth~\cite{fusco,1_3,Gralka2019,cancer_sampling3}, to determine the effects of spatial growth on three classic population genetics quantities: (i) time to the most recent common ancestor, (ii) number of segregating sites in a sample and (iii) clone size distribution of a sample. We find that these quantities are completely determined by the growth properties of the populations, and that their key features can be captured by a deterministic tree structure defined completely by the exponents of the KPZ universality class. 
We finally discuss how recent advances in lineage tracing \cite{2_main,2_1,2_2,2_5,2_10,2_13,1_12}, as well as single cell sequencing can be combined with our model to reveal the presence of surface-limited growth and interpret the data accordingly. 

%from tumors have become increasingly prevalent. This gives us an unprecedented amount of data to work with, which could be used to gain a more complete understanding of the underlying growth processes. In order to gain new insights from the data though, we still lack spatial growth models which could be used to analyze the data.

%\diana{Need better motivation and better contextualization}
%We will focus on constructing a model which can describe sampling data stemming from the outer parts of a population, rather than from the inner regions. This is due to the fact that in tumor sequencing, cells would typically be sampled from the outside due to the convenience of access of this region. Ideas on how to extend our model to also incorporate sequencing from inner layers will be given in section \ref{section4.3}.

%-----------------------------------------------------------------------------------------%
%-----------------------------------------------------------------------------------------%

\section{Simulating spatial growth: the Eden Model}
The Eden model, first introduced in the seminal paper by Eden in 1961 \cite{eden}, is widely used to mimic spatial growth processes where replication is limited to the front of the expansion, for example microbial colonies on rich media. Starting from an initial set of cells, placed at fixed points on a lattice, one cell with at least one empty neighbour is randomly chosen and replicated into one of the empty neighboring sites. This new cell can be seen as the descendant of the initially chosen cell and the process repeated to reach a final population size. 

The growth process simulated by the model can be tracked to generate a genealogical tree that identifies the mother-daughter relationship of each individual (lattice site). The statistical properties of the emerging lineages have recently been investigated~\cite{rand_walker} and found to fall within the KPZ universality class. 

This underlying growth process and the resulting lineages are sufficient to completely characterise the neutral genetic diversity of the population, since neutral mutations do not affect growth. The occurrence of neutral distinct mutations can be modeled as a Poisson stochastic process occurring on top of the identified lineages (infinite site model~\cite{Ewens2004}). Using the statistical properties of the lineages, we can then characterise the corresponding coalescence process and estimate classic population genetics quantities.
In what follows, we consider two scenarios: a linear front of constant width, which we compare to a Wright-Fisher model of constant population size; and a 2D radial expansion, which mimics colony growth. Details of the Eden model simulations are provided in Appendix~\ref{sec:sims}.

\section{Statistical properties of Eden model in a linear geometry} \label{sec:eden-linear}

We will start our analysis with a linear front scenario (corridor) so that the front of the population exhibits a constant width. In what follows, we will always sample individuals from the very front of the population.

\subsection{Number of segregating sites between two individuals}
Single-cell sequencing enables genomic comparison (either whole-genome or targeted regions) among individual cells sampled from different locations in the population. The number of differences between the two (or more) genomes is a well-studied summary statistics in population genetics called \textit{number of segregating sites}, $S$, whose distribution is known for well-mixed populations and even for simple models of structured populations (island model)~\cite{Ewens2004}. 

In contrast to the well-mixed scenario, the spatial structure of our model naturally raises the question of how the number of segregating sites $S$ depends on the relative location of the sampled individuals. Starting with two individuals sampled at a distance $d$ from each other, then the probability $P_2(S|d)$ of observing $S$ segregating sites is
\begin{equation}
    P_2(S|d)=2\int P_1(S|T)P(T|d)dT,
\end{equation}
where $P_1(S|T)$ is the conditional probability of observing $S$ segregating sites given that the time to the most recent common ancestor (MRCA) between the two individuals is $T$, $P(T|d)$ is the conditional probability of observing a time to the MRCA equal to $T$ given that the two individuals are sampled a distance $d$ apart. The factor 2 takes into account that mutations distinguishing the two individuals can occur on either branch leading to the MRCA. If we make the standard assumption that mutations follow a Poisson process, then $P_1(S|T)=\exp(-\mu T)(\mu T)^S/S!$, where $\mu$ is the mutation rate per replication and $T$ is in units of replication events. In the Eden model (see Appendix A), a replication event corresponds to the colonization of a neighboring lattice site, consequently both distances and times can be expressed in units of lattice sites.

\begin{figure*}[tb]
    \centering
    \includegraphics{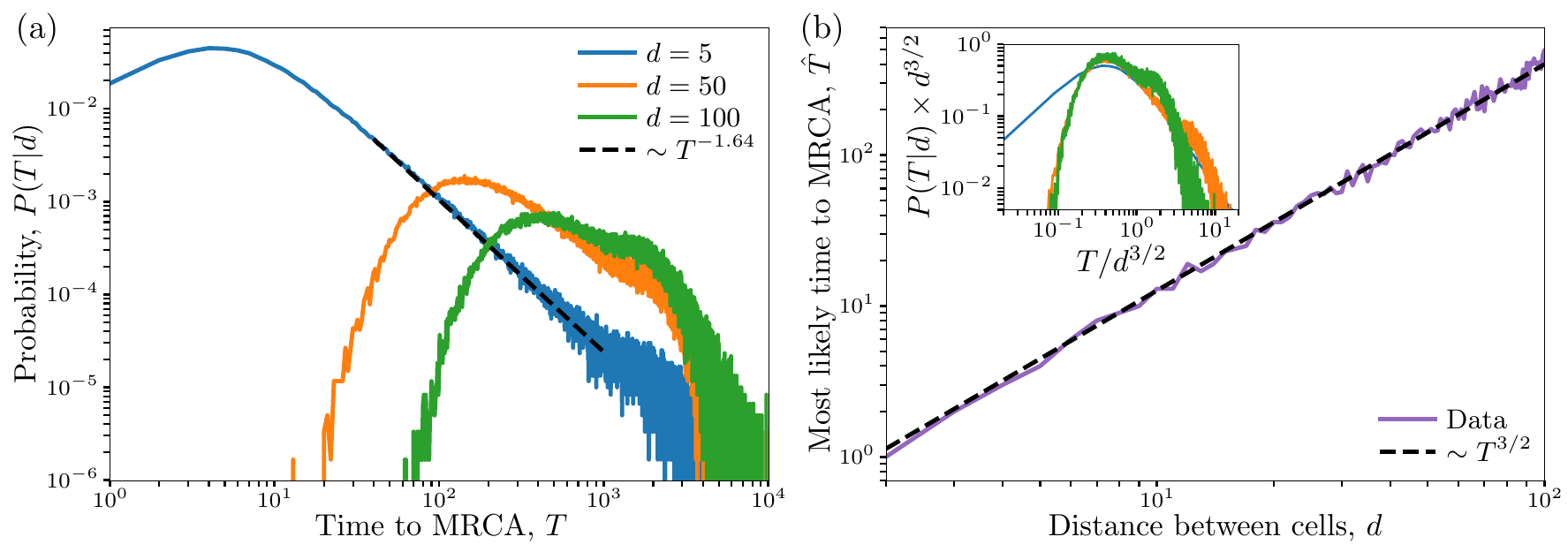}
    \caption{Left, Probability distribution of time $T$ to the most recent common ancestor for individuals sampled a distance $d$ apart, for simulations in a linear corridor. Right, dependence of the most probable value $\That$ of the distribution on the separation $d$. Inset shows the data on left, rescaled by the measured dependence of the peak of the distribution $\That \sim d^{3/2}$. The shown plots were each generated by averaging over the results of 600.000 simulation runs.}
    \label{fig:time_distribution}
\end{figure*}

The distribution $P(T|d)$ from linear simulations is shown in fig.~\ref{fig:time_distribution}, with a power-law decay at large distances following $P(T|d)\sim T^{-1.64}$. The exponent is connected to one of the characteristic exponents describing the statistics of directed polymers in random media (DPRM)~\cite{rand_walker}. As $d$ increases, simulations deviate from the power-law expectation at large times due to the finite size of the simulations. 

Because of the heavy tail of the distribution, the average time to the MRCA is often not of practical use, and the typical time to the MRCA, $\That$, is better suited as a metric of the characteristic behavior. This characteristic time has a scaling determined by the DPRM wandering exponent, $\That \sim d^{3/2}$ (fig.~\ref{fig:time_distribution}(b))~\cite{rand_walker}.

Since the probability function $P(T|d)$ decays quickly upon moving away from the characteristic value $\That$, a simplified model of the tree structure can be built by replacing the distribution with a $\delta$-function peaked at $\That$. In addition, for large $\mu T$, we can also approximate $P_1(S|T)$ to a $\delta$-function peaked at the mean value $\mu T$, so the distribution of segregating sites scales similarly to $P(T|d)$, rescaled by a factor $2\mu$. In particular, the most likely number of segregating sites observed $\hat{S}\approx2\mu\That\propto d^{3/2}$, following the KPZ expectation (fig.~\ref{fig:time_distribution}). For comparison, in a Wright-Fisher model of constant population size $N$, the typical time to the MRCA would be $N$, independently on the physical distance between the two sampled individuals~\cite{Ewens2004}. If we equate the population size $N$ to approximately the width $w$ of the corridor, this leads to a critical distance $d^*=N^{2/3}<w$, so that if individuals are sampled at distance larger than this, they should show more segregating sites than the well-mixed expectation, and viceversa.

% \begin{figure}[h]
%     \centering
%     \includegraphics[width=0.5\textwidth]{mrca.png}
%     \caption{CAPTION CAPTION CAPTION CAPTION CAPTION CAPTION CAPTION CAPTION CAPTION CAPTION CAPTION CAPTION CAPTION CAPTION CAPTION CAPTION }
%     \label{fig:mrca}
% \end{figure}

\subsection{Number of segregating sites in a subsample}

If we consider a  connected subsample $n<N$ at the front of the population, the total number of segregating sites is related to the total length of all the branches in the genealogical tree $T_\text{tot}(n)$. In the Wright-Fisher model this leads to the well-know average result
\begin{equation}
    T_\text{tot}=\sum_{i=2}^{i=n}iT_i=2N\log(n-1),
\end{equation}
for an haploid population, where $T_i=N/{i\choose 2}$ represents the average time for the first coalescent event between two lineages of the possible $i$~\cite{Wakeley2008}. Because after each coalescence event the number of lineages decreases by one, the total length of the tree is just given by the sum of the number of surviving lineages between subsequent coalescent events. Note that the time to the MRCA across $n$ individuals in a well-mixed population of size $N$ is $\Tmrca(n)=\sum_{i=2}^{i=n}T_i=2N[1-1/(n-1)]$.

\begin{figure}[t]
  \centering
  \includegraphics{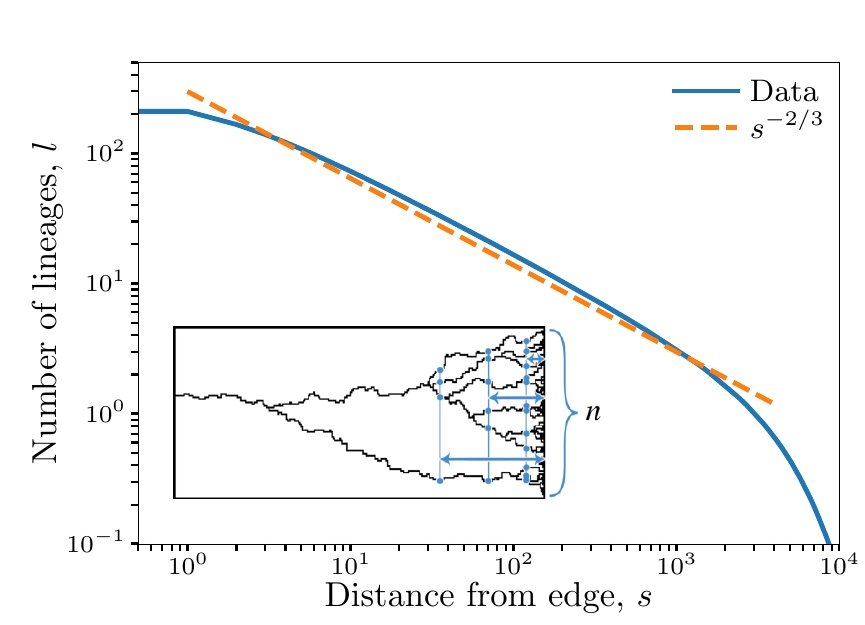}
  \caption{Number of surviving lineages in the genealogy of $n$ individuals at
    the edge of the colony, traced backwards as a function of the distance $s$
    from the edge of the colony. Inset shows an example of a genealogy from an
    Eden model simulation. The shown plot was generated by averaging over the results of 120.000 simulation runs.}
  \label{fig:lineages_corridor}
\end{figure}

For the Eden model, the expression depends on the relative position of the $n$ individuals. If we assume that they are positioned contiguously along the front, then $T_\text{tot}$ is the total length of the branches that lead to a corridor of width $n$ starting from the MRCA of the $n$ individuals (as in the inset of fig.~\ref{fig:lineages_corridor}). Then
\begin{equation}
    T_\text{tot}=\int_1^{\Tmrca(n)}l(t)dt,
\end{equation}
where $l(t)$ is the number of lineages at time $t$ measured backwards from the subsample and $\Tmrca(n)$ is the time to the MRCA of the whole sample. Because of the spatial constraints on the lineages, we have that $\Tmrca(n)\propto n^{3/2}$, since the $n$ individuals will be at most $n$ lattice cites apart (in reality there are more than $n$ individuals in a width $n$ since the front is rough, but we use this as first approximation). 

The scaling of the typical time to the MRCA implies a particular scaling for the number of lineages with reverse time. We assume that all coalescence events happen exactly at the typical time $\That(d)\sim d^{3/2}$ associated with the separation $d$ of two individuals on the front. Upon advancing backwards in time from the front by an interval $t$, each contiguous segment of the boundary of size $\sim t^{2/3}$ will have coalesced to a single ancestor. As a result, the number of surviving lineages $l(t)$ falls as $t^{-2/3}$ with reverse time $t$, as long as the number of surviving lineages is large. To test this relation, we measured the number of lineages as a function of distance $s\propto t$ from the edge of the colony in Eden simulations within a linear corridor (see \figref{fig:lineages_corridor}). Although the true genealogies display a more complex structure compared to our idealized model, we find that the observed behaviour in $l$ is consistent with a power-law decay $l \sim s^{-2/3} \sim t^{-2/3}$ until only a few lineages survive ($l \lesssim 1$). % the last two lineages coalesce (the average $l(t)$ is 1). This result can be interpreted analytically as if each lineage, going backwards in time, was following a super-diffusive random walk and annihilating with another lineage when they meet \diana{maybe we can give a bit more details here}.

Since the number of lineages is $l(t)\sim n/t^{2/3}$, then
\begin{equation}
    T_\text{tot}\sim n\Tmrca^{1/3}(n)\propto n^{3/2}.
\end{equation} 
A remarkable byproduct of this is that the total tree size is proportional to the $\Tmrca(n)$ of the sample, unlike the case for well-mixed populations, and it does not depend on the total population size. This reflects the fact that the dominant contribution to the tree is given by the significantly longer oldest branches. %, and it also suggests that trees with different $n$ leaves (as long as they are contiguous) are self-similar.
Another interesting feature arising from the comparison with the Wright-Fisher model is that the total number of segregating sites increases much more quickly with sample size for the spatial model than for the well-mixed one. %Comparison between this analytical expectation and the simulations results are in fig.... \diana{It would be good to compare this theoretical prediction with the simulations.} \jayson{I'm a bit confused by this. Wouldn't geometry dictate that $T_{mrca} = T_\text{tot}$ because they encompass the same 'triangle' for contiguous sites? It could be that I don't understand $T_\text{tot}$ correctly.} \diana{The notation is probably bad. $T_{mrca}(n)$ is the time to the most recent common ancestor of $n$ leaves. $T_\text{tot}(n)$ is the total tree length (sum of all the branches) that start from the MRCA and lead to the $n$ final leaves. Thinking about it, the fact that they are proportional to each other (that's not the case for well-mixed) implied the kind of renormalization group observations that Armin made graphically looking at this tree. I will highlight this. }

\subsection{Site frequency spectrum of subsamples} 
The statistical properties of the number of segregating sites in a sample $n$ of the edge of the population also determine the mutational spectrum, a commonly used genomic metric of the population structure. In the case of the Wright-Fisher infinite site model, the mutational spectrum, i.e., the number of mutations $m(j)$ carried by $j<N$ individuals, is given, on average, by $m(j)=2\mu N/j$~\cite{Ewens2004}.  

The hierarchical length structure of the genealogical tree generated by the Eden model generates a very different mutational spectrum, since mutations can accumulate for a long time on long lineages before any later branching event occurs. Importantly, to understand the origin of this mutation spectrum, the topology of the tree (which branches coalesce with each other) is crucial, as identical $l(t)$ can generate very different mutational spectra. %\diana{(see fig. add cartoon in SI, maybe?)}.
Furthermore, the measured spectrum will depend strongly upon the spatial distribution of the population samples---different sampling protocols might be sensitive to different features of the genealogical structure.

If samples are taken uniformly across the entire population, the site frequency spectrum is expected to follow a trend $m(j)\propto j^{-7/5}$~\cite{fusco}. However, in many situations the outer edge of the population is more accessible for sampling. An analogous theoretical argument can be made if we restrict sampling to the edge of the population. Then, mutations carried by at least $j$ individuals have to occur before the coalescence time $\Tmrca(j)$ or, in other words, somewhere on the subtree between the ancestor among all $N$ individuals and the ancestor of the subsampled $j$ individuals. We will call this portion of the total tree $T^*(j)$. The topology of the tree then determines the expression for the site frequency spectrum. 

For instance, if the tree is well-balanced so that coalescence events happen almost at the same time between pairs of lineages that are the same distance apart, then the number of mutations $m(j)$ carried by at least $j$ individuals is proportional to
\begin{equation}
T^*(j)=T_\text{tot}(N)-T_\text{tot}(j)\frac{N}{j},
\end{equation}
as there would be $N/j$ identical subtrees emerging from the corresponding $N/j$ lineages, each with $j$ leaves on average. This leads to
\begin{equation}
    T^*(j)\approx N^{3/2}- j^{3/2}\frac{N}{j}=N^{3/2}(1-x^{1/2}),
\end{equation}
where $x=j/N$ is the frequency of the mutation. % [Note, one reaches the same result by integrating the number of lineages between $T_{mrca}(n)$ and $T_{mrca}(j)$.] 
This expression leads to a clone size distribution $\Pi'(x)$ (probability that a mutation is carried by a proportion $x$ of the population) to be proportional to $x^{-1/2}$.%, as $m(x)\propto\frac{d(1-T^*(j))}{dx}$. 

\begin{figure}
  \centering
  \includegraphics{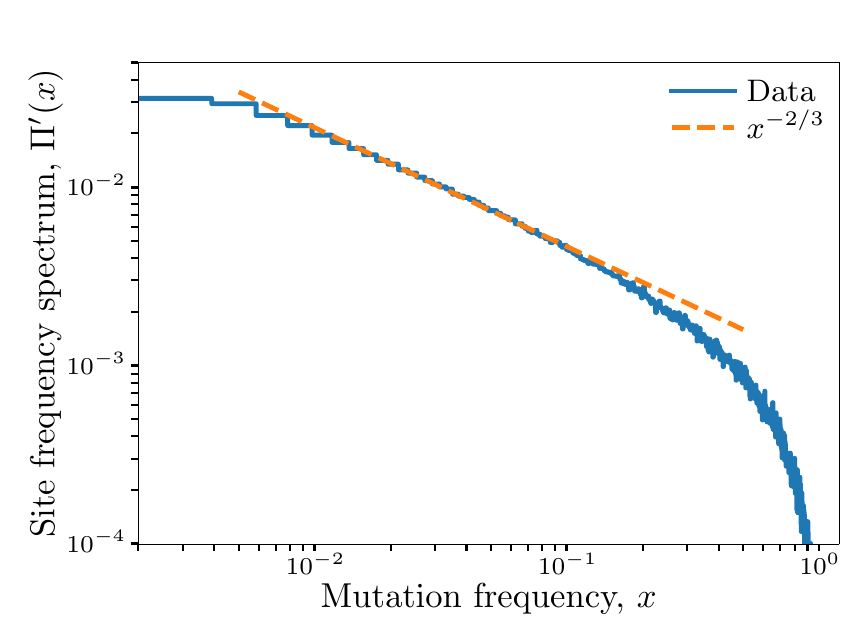}
  \caption{Site frequency spectrum, i.e. probability of a mutation attaining a frequency $x$, for individuals sampled from the edge of a population generated using Eden model simulations in a corridor geometry. The shown plot was generated by averaging over the results of 24.000 simulation runs.}
    \label{fig:sfs-corr}
\end{figure}

Eden model simulations show that, at least for small $x$ when finite size effects are limited, the scaling is $\Pi'(x)\propto x^{-2/3}$ (\figref{fig:sfs-corr}), clearly indicating that the tree, in this case, is not balanced, and the tail of the $\Tmrca$ distribution plays a crucial role. Importantly, the resulting clone size distribution is less steep than the well-mixed scenario, which corresponds to a higher likelihood of finding a mutations carried by a large proportion of the sample, compared to a well-mixed population. 

%Also, because the older branches contribute the most to the tree size, it's more likely that exactly the same $k$ mutations will be carried by the same individuals.

%(I THINK THE EXTRA N/J DEPENDS ON THE TOPOLOGY OF THE TREE, BUT WE CAN COMPARE WITH THE RESULTS. IT DOES MAKE A BIG DIFFERENCE. If for instance one assumes that it branches one at a time, you get $(1-x^{3/2})$). This calculation assumes that topology of the tree is symmetrically balanced, so that subgroups of $j$ individuals among the total $N$ will coalesce with their neighbors around the same time. Fig.. shows a comparison of the different theoretical expectations and the simulation results.  

%\subsection{Accumulation of mutations}
%A third summary statistics that can easily be assessed via single cell sequencing is the number of mutations accumulated during growth by an individual. If we condition on the total number of replication events, the individuals at the end of the simulation or at the edge of the Eden model will have approximately undergone the same number of generations, so we expect that the typical number of mutations accumulated by each individual should be similar in the two models and proportional to the number of generations. However, because the site and allele frequency spectrum is going to vary dramatically, also the probability of carrying \textit{specific} combinations of mutations.

%\subsection{Population subsampling}

\section{Radial expansion}
While linear expansions are useful to understand the properties of the tree structure generated by the spatial growth process, radial expansions are more relevant to several real case scenarios (e.g., microbial colonies, avascular tumors). In this case, the population expands initially very rapidly due to an inflation effect related to curvature, which slows down as the radius grows~\cite{Korolev2010}. 

\begin{figure}[h]
    \centering
    \includegraphics{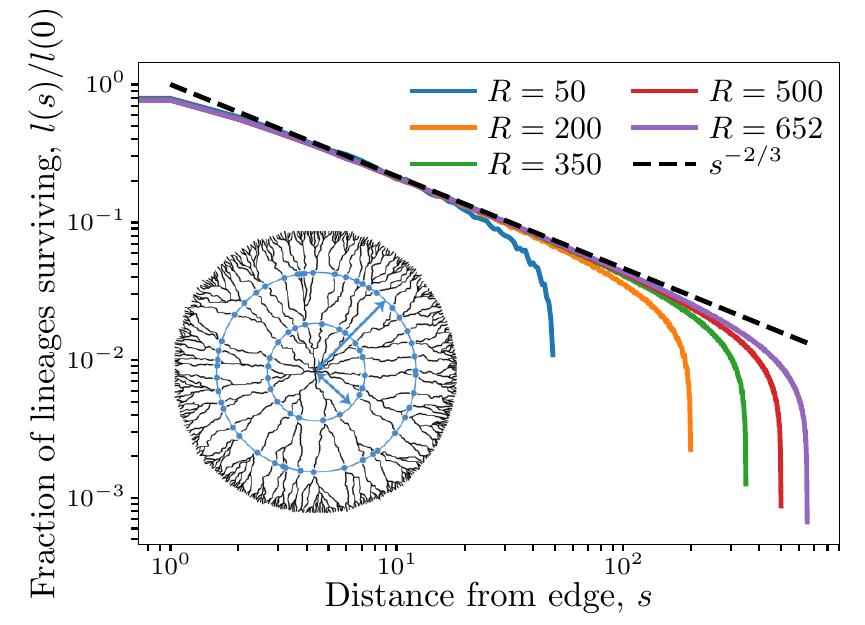}
    \caption{Fraction of lineages which survive when the genealogy of all individuals at the edge of the colony is traced inward, as a function of the distance $s$ from the edge of the colony. Solid lines show data from radial Eden simulations grown to different final radii $R$. Dashed line shows expected decay based on lineage fluctuations in the KPZ class. Inset shows an example of a genealogy for a radial Eden simulation. The shown plots were each generated by averaging over the results of 48.000 simulation runs.}
    \label{fig:rad2lin}
\end{figure}

From radial Eden model simulations, we find that the number of lineages as a function of the distance $s$ from the edge of the colony follows the same $s^{-2/3}$ power-law as in the corridor case, with a sharp drop close to the centre of the colony where the lineages spread star-like due to the rapid inflation process (fig.~\ref{fig:rad2lin}). 

%One can now re-scale the horizontal axis to display the fractions of the total colony radius and let the vertical axis display the number of lineages divided by the circumference of a circle of radius equal to the specific distance away from the center, i.e. the vertical axis displays the lineage density. This leads to the plot shown in the top inlay of figure \ref{fig:collapsed}. 
\begin{figure}[h]
    \centering
    \includegraphics{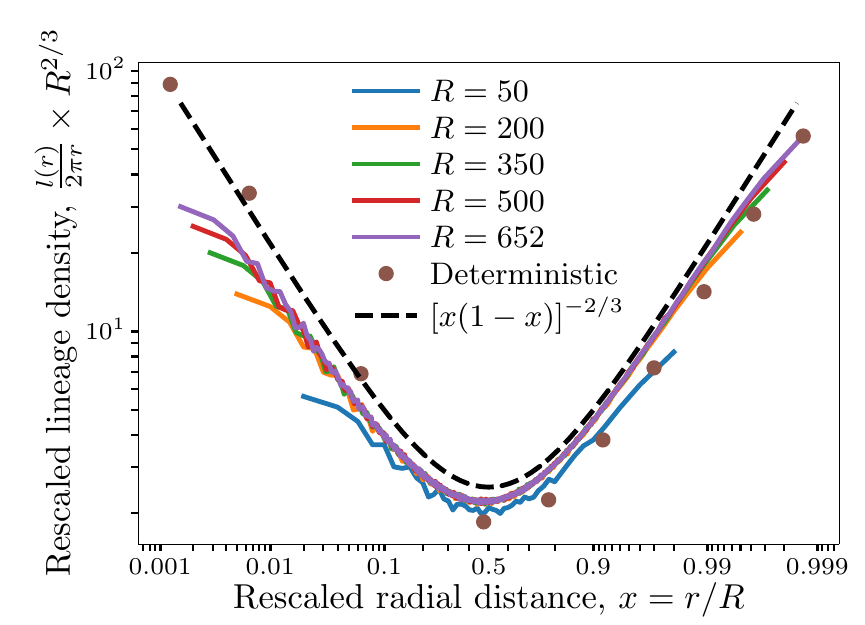}
    \caption{Lineage density as a function of distance from the colony center, rescaled by the final colony radius. The horizontal axis is a logit scale which reveals the power-law divergences in the lineage density as $r/R \to 0$ and $r/R \to 1$. Solid lines are measured from Eden simulations at different final sizes $R$; symbols are from the universal tree model with $p = 4$; dashed line is a phenomenological master curve (\eqnref{eq:density_th} with $\epsilon=1$). The shown plots were each generated by averaging over the results of 48.000 simulation runs.
    }
    \label{fig:collapsed}
\end{figure}

To account for the inflation process, we normalize the number of lineages present at radius $r$ by the circumference of the colony at the same radius to obtain a lineage density, and rescale the radius $r$ by the final colony radius $R$ (fig.~\ref{fig:collapsed}). We observe that consistently across colony sizes, the lineage density undergoes a transition at exactly a radius $R/2$, where $R$ is the radius of the final colony, so that the density of surviving lineages initially decreases and then increases with $r$. This non-monotonic behaviour reflects the tradeoff between the process of inflation, which pushes lineages apart preventing them from coalescing, and the stochastic wandering of the lineages, which over time makes them coalesce. %As shown previously in radial expansions (cite), the mean square displacement of a lineage (random walker) that has survived until radius $r$ can be transformed to a mean square angular displacement $\langle\Delta\alpha^2\rangle$ that satisfies the following scaling:
%\begin{equation}
 %   \langle\Delta\alpha^2(\Delta r)\rangle\sim\frac{\Delta r^{2\zeta}}{(r+\Delta r)^2},
%\end{equation}
%where $\Delta r$ is the distance travelled by the lineage. Importantly, independently of the type of diffusion (coefficient $\zeta$), this function has a maximum at $\Delta r=r$. \diana{this is from Oskar's notes, but it's not clear to me how relevant it is!}

The collapse of the lineage density on a master curve independently of colony size suggests the presence of a universal tree that can describe the behaviour of genealogies generated in these two-dimensional spatial growth models. We propose it below. 

\subsection{\label{sec:section2}Universal tree model}
\begin{figure}[h]
    \centering
    \includegraphics[width=0.35\textwidth]{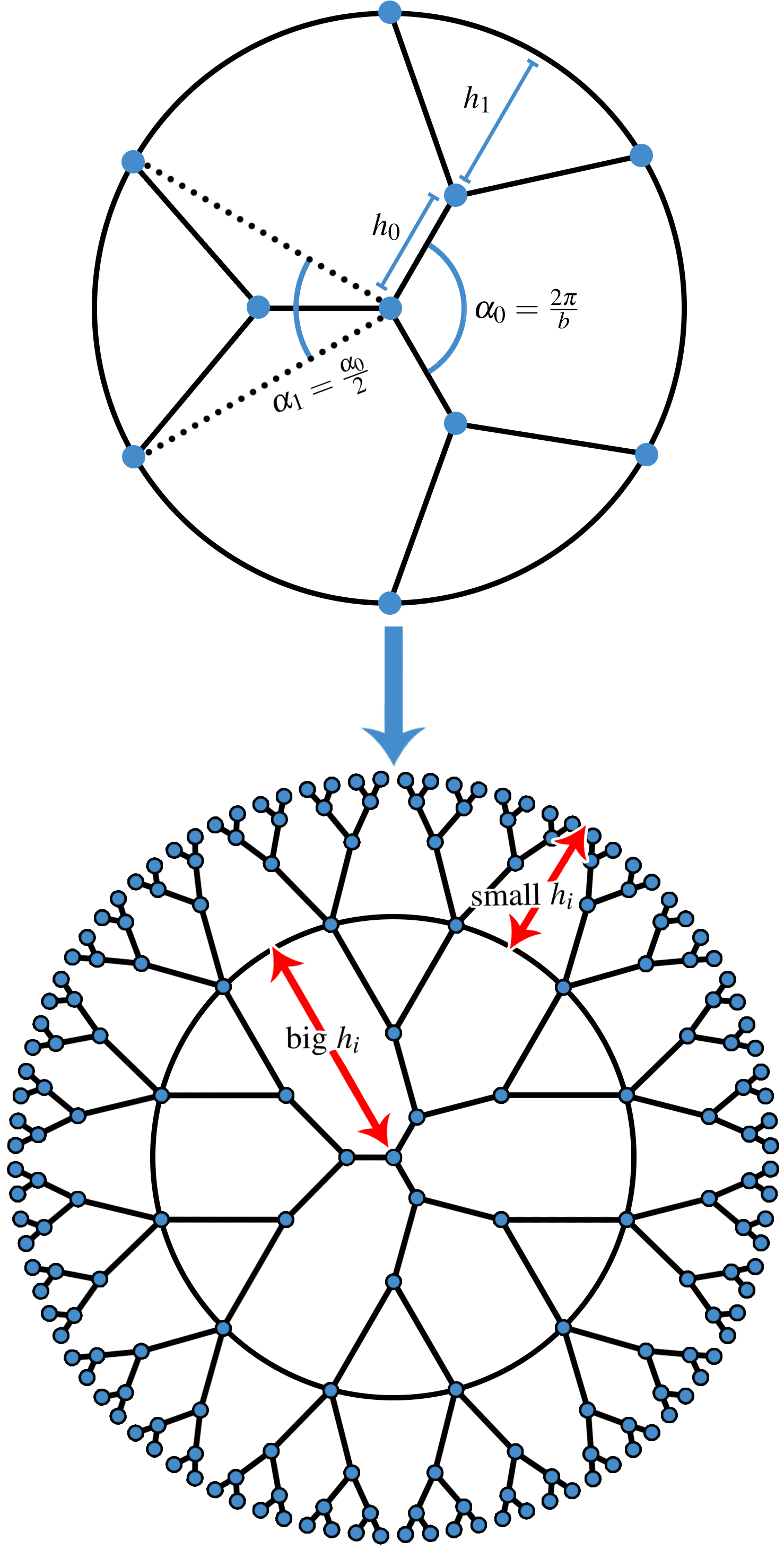}
    \caption{Sketch of the deterministic tree model that captures that average lineage properties during radial growth.}
    \label{fig:model}
\end{figure}

Our proposed model incorporates the statistics of lineage fluctuations imposed by the KPZ universality class, which the Eden model is known to belong to, as well as the spatial constraints on coalescence arising from the radial structure of the expansion (\figref{fig:model}).
%Our proposed model is based upon the KPZ universality class \diana{what does kpz stand for? Citations?}, which describes bacterial colony growth.\diana{In the intro you were talking about tumors, now about colony growth. Why is this a reasonable model to start with?} From the scaling exponents of the KPZ class we derived a relation between the lateral shift of cell lineages and the lineage's growing time\diana{what are these quantities? Need a sketch to describe what you are talking about}: 
%\begin{equation}
 %   \textrm{lineage's lateral shift}\sim t^{1+\frac{\alpha-1}{z}} = t^{2/3},
%\end{equation}
%where $t$ is the growing time of the lineage and $\alpha = 1/2$, $z = 3/2$ are the KPZ universality class's scaling exponents. This derivation was inspired by \cite{hallatschek_eden}, where a similar result for the boundaries between growing populations was derived. Based on this relation, 
Building on our results for the corridor case, we hypothesize that the lineages of two cells which lie a spatial arc distance $d$ apart at a radius $r\gg d$ from the center of the colony will most likely coalesce at a certain distance $h$ towards the center of the colony, so that
\begin{equation}\label{eq:3/2}
h \sim d^{3/2}.
\end{equation}

We will now neglect rare stochastic events in the coalescence process of the colony's lineages and devise a deterministic model where coalescence is controlled only by the typical coalescence height. The model is based on a binary tree with its branches' lengths following the above relation \ref{eq:3/2}. We will also assume that always exactly two lineages coalesce into one and that all lineages in the tree that are a distance $d$ apart coalesce at the same time (perfectly balanced tree, fig~\ref{fig:model}).

An important difference between the corridor and the radial case is that in the radial case the MRCA of the whole population is always clearly identifiable. The symmetry relation in lineage densities highlighted in fig.~\ref{fig:collapsed} then allows to build our tree forward in time from the centre of the colony as a branching process (rather than backwards as we did for the corridor). 

We start with an arbitrary number $b$ of starting branches, growing star-like towards the outside, splitting at deterministic distances away from the center based on the distance for which the branches have already been running, according to eq. \ref{eq:3/2}. %\diana{Since you are growing the tree inside-out, I wouldn't call this coalescence, but branching events (forward in time)} 
Since we are in a radial setting with an inflation term, the distances between the lineages is dictated by both the angle between them, as well as the current distance away from the center. %In figure \ref{fig:model} one can see a sketch of how the model tree is constructed. \diana{need to make this figure nicer (probably from Armin's thesis}

After each splitting step, the angles are halved, leading to 
\begin{equation}\label{eq:alpha_i}
   \alpha_i = \frac{\alpha_{\textrm{start}}}{2^i}\textrm{ for }i\in \{0,1,2,\dots\} \textrm{, with } \alpha_{\textrm{start}} = \frac{2\pi}{b}.
\end{equation}
For the very first step ($i=0$), where $b$ branches start from the center, the height to splitting $h_0$ is calculated as 
\begin{equation}\label{eq:h_0}
\begin{split}
h_0 &\sim (\alpha_0 h_0)^{3/2}\\
\implies h_0^2 &\sim \alpha_0^3h_0^3\\
\implies h_0 &\sim \frac{1}{\alpha_0^3},
\end{split}
\end{equation}
where we used relation \ref{eq:3/2} together with the fact that in the very first step, the distance until the next splitting point is the same as the distance away from the center used for calculating the arc length distance between the lineages. For all following steps, the splitting heights are given via the equation 
\begin{equation}\label{eq:h_i}
    h_i \sim\Bigg( \alpha_i \bigg( \sum_{j=0}^{i-1}{h_j} + h_i\bigg) \Bigg)^{3/2}\textrm{ for }i\in\{1,2,3,\dots\}.
\end{equation}

For all $i>1$, equation $\ref{eq:h_i}$ can be solved numerically. One finds that the equation always gives two positive solutions. As $i$ increases, the next $h_i$ is always dependant on the result of the previous heights $h_0,\dots ,h_{i-1}$, requiring some considerations on the choice of solution. We find that choosing the bigger solution leads to increasing values of $h_i$ until they diverge. Reciprocally, choosing the smaller solution leads to decreasing $h_i$ that converge towards $0$. It is important to point out here that, while we use the scaling in equation \ref{eq:3/2} to determine the location of the branching events, the relationship is expected to hold only when the distance between lineages is much smaller than the radius of the branching event ($d\ll r$). For the first few steps this assumption is likely to break. However, since lineages double at every step while the time between steps grows at best as a power-law, the condition becomes true relatively quickly in the expansion process ($i>2$).

The simulation results for lineage density suggest that the distance between branching events becomes increasingly longer in the first half of the colony growth (up to $R/2$) after which it then becomes shorter and shorter with each subsequent step. As a result, for our model, we will choose the larger solution for the first $p$ steps, and subsequently always choose the smaller solution. In the abstract model, the branching process can continue indefinitely, creating ever-shorter branches spaced closer and closer together as the tree grows outward. However, the successive values of $h_i$ when $i > p$ decline so rapidly that the sum $\sum_{i=0}^\infty h_i$ converges to a finite value which corresponds to the radius $R$ of the colony. Upon assembling deterministic trees with different values of $p$, we observe that $\sum_{i=0}^p h_i \approx \sum_{i=p+1}^\infty h_i \approx R/2$ independently of the value of $p$. The geometry dictated by the solutions to equation \ref{eq:h_i} ensures that the transition from choosing the larger to the smaller solution always happens at approximately half of the colony radius, which coincides with the  lineage density minimum in the Eden simulations (fig.~\ref{fig:collapsed}).

The universal tree model provides a deterministic prediction for the lineage density. As \figref{fig:model} shows, the number of lineages doubles at discrete values of the distance from the central node. By recording these values and the corresponding number of branches, we obtain a rescaled lineage density (symbols in Fig~\ref{fig:collapsed}) which reproduces the curves measured from Eden simulations. The behavior as $x = r/R \to 1$ is dictated by the power law $l(s) \sim s^{-2/3}$ expected from the corridor geometry and confirmed in Fig.~\ref{fig:rad2lin}: upon using the relation $s = R(1-x)$, we have $l(r)/(2 \pi r) \sim l(r)/(2 \pi R) \sim [R(1-x)]^{-2/3}$. The universal tree recovers this scaling at large $x$, but also reveals the behavior of the lineage density for $r \lesssim R/2$, where inflation and branching play opposite roles (inflation creates space for lineages and branching events quickly fill it up). We find that the number of branches in the universal tree grows as $l(r) \sim r^{1/3}$, leading to a divergence $l(r)/(2 \pi r) \sim r^{-2/3}$ in the lineage density as $r \to 0$. These two asymptotic behaviors are well-captured by a phenomenological master curve
\begin{equation}
  \label{eq:density_th}
  \frac{l(r)}{2\pi r} = \frac{1}{\epsilon^{1/3}R^{2/3}} \left[x(1-x)\right]^{-2/3},
\end{equation}
where $\epsilon$ is a small-distance cutoff equal to the lattice spacing in the Eden model simulations (see Appendix \ref{sec:cutoff} for details). Equation \eqref{eq:density_th} (dashed line in \figref{fig:collapsed}) successfully reproduces the lineage densities measured in Eden model simulations as well as the deterministic geometry generated by the universal tree model.

\subsection{MRCA position of contiguous subsamples}

% Fig.~\ref{fig:rad2lin} shows that if the MRCA is sufficiently close to the edge (so for small $n$), the number of lineages decays similar to the corridor case, as the distance between the lineages is much smaller than the radius of the colony, so any curvature effect can be neglected. However, this relationship is expected to fail as the $\Rmrca$ approaches $R/2$ (large sample sizes).
The proposed master curve for the lineage density, \eqnref{eq:density_th}, combines the effects of inflation and stochastic coalescence in a succinct form which explicitly captures the distinct tree structures for center distances below and above the value $r = R/2$.
As an application of our results to a quantity of relevance to typical biological measurements, we now use this master curve to determine a general relationship for the radial distance $\Rmrca$ of a contiguous sample of size $n$ taken at the outer boundary of a colony that has reached a final radius $R$.

We measure distances and sample sizes in units of the lattice spacing, and correspondingly set $\epsilon = 1$ in \eqnref{eq:density_th}. 
If we assume a uniform angular distribution of lineages, then the number of surviving lineages for a contiguous sample of size $n$ (covering an angle $n/R$) varies with distance $r$ from the colony centre as
\begin{equation}
    l(n,r)=l(r)\frac{n}{2\pi R}.
\end{equation}
Then the MRCA corresponds to the radius at which we are left with only one lineage $l(n,\Rmrca)=1$, leading to the following equation
\begin{equation}
    x^{1/3}(1-x)^{-2/3}=R^{2/3}n^{-1},
\end{equation}
which can be solved exactly. Because $0<x=\Rmrca/R<1$, the acceptable solution to this equation is always unique and corresponds to
\begin{equation}
    x_\text{MRCA}=\frac{1+2 R^2 n^{-3}-\sqrt{1+4 R^2 n^{-3}}}{2 R^2 n^{-3}}.
\end{equation}

For large $n$, $\Rmrca<R/2$ and the scaling with sample size is such that $\Rmrca \sim n^{-3}$. Conversely, if $n$ is small, the MRCA is close to the edge on average, and its position follows the scaling $1- \Rmrca/R\sim n^{3/2}$, which is analogous to the corridor case. 

The transition between the two regimes corresponds to when $\Rmrca\approx R/2$. Using the equation above, we find that this corresponds to a critical angle (and critical sample size) $\alpha_c=n_c/R\approx R^{-1/3}$. This scaling has been previously identified as the threshold frequency between bubbles and sectors in neutral mutations in two dimensional colonies~\cite{fusco}.

\subsection{Number of segregating sites}

Analogously to the corridor case, the number of segragating sites $S$ in a sample of size $n$ is proportional to the total tree size that leads to the $n$ surviving leaves, so that
\begin{eqnarray}
     T_\text{tot}(n)&=&\int_{\Rmrca}^{R}l(n,r)dr\\&\propto& n R^{1/3} \int_{\Rmrca/R}^1 x^{1/3}(1-x)^{-2/3}dx\\
     &=&nR^{1/3}\left[\mathcal{B}\left(\frac{4}{3},\frac{1}{3}\right)-\mathcal{B}\left(x_\text{MRCA}(n),\frac{4}{3},\frac{1}{3}\right)\right]
\end{eqnarray}
where $l(n,r)$ are the number of lineages as a function of the distance from the center $r$ that lead to the sample and $\mathcal{B}$ indicates the corresponding beta function.
 
\subsection{Site frequency spectrum}
Our tree model assumes a perfectly balanced tree with $b$ initial branches that set the largest possible frequency of a mutation in the front population ($1/b$). Similarly to the argument for the corridor case, a mutation that is carried by at least $n$ individuals at the edge has to occur somewhere in the tree before $\Rmrca(n)$. Then, the number of mutations $m(n)$ carried by at least $n$ individuals at the edge is proportional to 
\begin{eqnarray}
    T^*(n)&\propto&\int_0^{\Rmrca(n)}l(r)dr\\&\propto& R^{4/3}\mathcal{B}\left(x_\text{MRCA}(n),\frac{4}{3},\frac{1}{3}\right).
\end{eqnarray}

\begin{figure}[h]
    \centering
    \includegraphics{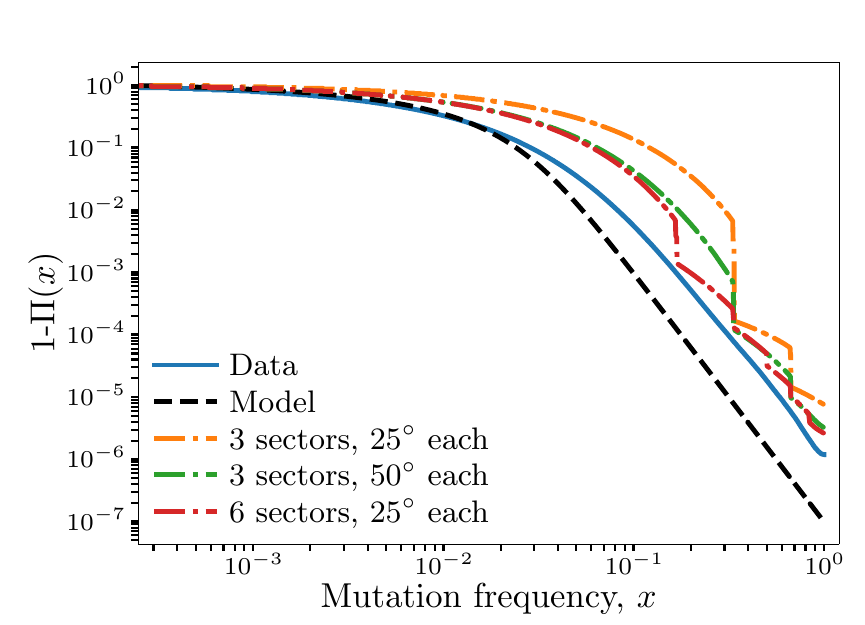}
    \caption{Comparison in the cumulative site frequency spectrum from simulations sampling from the whole edge (solid blue line), equally spaced subsamples of different size and number (dash-dotted lines) and the theoretical expectation from eq.~19 (black dashed line). No fitting parameters are necessary. The shown subsampling plots were each generated by averaging over the results of 18.000 simulation runs, while the whole edge plot averaged over 48.000 runs.}
    \label{fig:sfs-radial}
\end{figure}

Fig.~\ref{fig:sfs-radial} shows that the theoretical expectation of the cumulative site frequency spectrum ($1-\Pi(x)$, i.e., probability that at a mutation is carried by at least a fraction $x=n/(2\pi R)$ of the population), without any additional fitting parameter, agrees remarkably well with the simulation. We observe a slight deviation at the point of inflection due to the discretized nature of the lattice in the simulations (the edge is only approximately one site thick). Interestingly, the agreement between theory and simulations suggests that in the radial case the tree is much more balanced then in the corridor case. The power-law tail, corresponding to an exponent of $-4$, is consistent with the site frequency spectrum of the full colony as we expect the large frequency mutations at the periphery to be stemming from sectors. The low frequency component of the site frequency spectrum is, in contrast, almost flat reflecting the fact that the later portion of the tree contributes negligibly to the total tree size.

In practice, often, only a subsample (or subsamples) of the colony periphery may be sequenced, as for instance in tumour biopsies~\cite{cancer_sampling1,cancer_sampling2, cancer_sampling3}. Because of the spatial correlation of the genealogies, these subsamples can exhibit unusual signatures in the site frequency spectrum.
We have found above that if two individuals (or samples) are picked farther than $\alpha_c$ apart, their MRCA will very quickly converge to the centre of the colony. This implies the presence of long independent lineages that lead to the different samples over which several mutations can accumulate. Because these lineages do not branch for a long time, they will lead to a large number of mutations carried by a very specific frequency in the sample, showing up as sudden drops in the cumulative site frequency spectrum (fig.~\ref{fig:sfs-radial}, dash-dotted lines). The position and size of these drops depends on the geometry of the sampling scheme.

If we have $N$ samples of size $n$ (each covering an angle $\alpha=n/R$), the $\Rmrca(n)$ of each sample is given by equation 15. From this point to the edge, we expect, on average, the $N$ trees to be similar and thus no mutation drop should be observed for frequencies below $1/N$. The topology of the tree for $r<\Rmrca(n)$ determines the position and size of the drops we observe in simulations. Since the number of leading lineages is $N$, in principle we can expect to see drops at any frequencies $i/N$ with $i
\in \{1..N-1\}$, each corresponding to the length of tree lineages shared by $i$ of the $N$ samples. The length of such lineages depend on the separation between samples.
As the sample size $n$ or the sample number $N$ increases, we expect the site frequency spectrum to converge to the full edge.

%\diana{I could do more precise calculations here, but for now I think it's good. We can do it in the revision.}

\section{Discussion and conclusions}

In this work we have analyzed the coalescence process generated by two-dimensional spatial growth models to provide quantitative expectations for some typical genetic observables that can easily be determined from population sequencing, such as the number of segregating sites and site frequency spectrum. Our analysis extends previous work on the topic first, by introducing an infinite site model on top of the growth process and thus going beyond the typical assumption of low mutation rate and second, by considering practical situations in which only a subset of the population is sampled. While here we focus on the 2D Eden model as a specific example of spatial growth that has been shown to well capture the statistical properties of microbial colonies, our analysis can be easily applied to three dimensional growth and to other types of random-walk models outside the KPZ universality class. In particular, 3D Eden model simulations have been shown to also display bubbles and sectors~\cite{fusco} analogously to the 2D case. We thus expect that a non-monotonic lineage density profile qualitatively similar to the one showed in fig.~\ref{fig:collapsed}, but characterised by different exponents, could be found also in this case, which is relevant to tumor growth. 

Our results show, in agreement with previous work~\cite{rand_walker}, that the lineages generated by an Eden model behave like directed polymers and can thus be modeled as random curves with super-diffusive statistics (mean-square transverse displacements grow faster than linearly with lineage length). The coalescence process (backwards in time) is then dictated by the annihilation of pairs of lineages as they collide. This analogy allows us to find a mathematical formulation for the average number of lineages as a function of time that lead to a final population at the edge of the expansion, which then can be used to provide estimates for the time to the MRCA and the number of segregating sites. Estimates for the site frequency spectrum require knowledge of the tree topology. Interestingly, here we find that the radial expansion is consistent with a balanced topology. By contrast, a linear front generates a site frequency spectrum inconsistent with a balanced topology, suggesting that rare long branches which coalesce well past the typical coalescence time play a crucial role. 

While in this work we use the Eden model to describe the growth dynamics of two-dimensional populations building on previous studies~\cite{hallatschek_eden,fusco}, this is by no means the only possible choice. The Eden model and the underlying KPZ universality class are relevant when the expanding front increases its roughness as it advances, due to a geometric feedback between local deviations from smoothness and the addition of new material at the front~\cite{kpz}. In some populations, however, front roughness might be suppressed and growth models that maintain a locally flat front are more appropriate.  Alternative lattice models which maintain flat fronts, such as the Domany-Kinzel model~\cite{Lavrentovich2013} or flat-front stepping stone model~\cite{Cox2002} have been used to study clone dynamics in range expansions where front roughening is absent. The main difference compared with KPZ-type models is that the transverse wandering of lineages follows diffusive statistics (wandering exponent of $1/2$)~\cite{rand_walker}, and can be described by a Langevin equation with multiplicative noise~\cite{Korolev2010,Lavrentovich2013}. Our approach, with appropriately modified exponents, could be used to construct simplified genealogies for flat-front models of colony growth.

Recent studies have used both deep sequencing and lineage tracing techniques to generate a vast amount of data to disentagle tumour growth dynamics and selection~\cite{1_12,1_14,cancer_sampling1,cancer_sampling2,cancer_sampling3}. In many cases, sector-like patterns are clearly observable suggesting that spatial growth and competition at the edge of the expansion play a crucial role.  Because the cellular population is not well-mixed, results obtained from local sampling need to be carefully interpreted to infer the dynamics at play. For instance, the accumulation of mutations at specific frequencies, which in a well-mixed scenario would be interpreted a signature of selection, can be a sole consequence of the geometry of the sampling scheme, making evolutionary inference particularly challenging~\cite{cancer_sampling3}. In this context, our analysis provides guidelines to design sampling schemes that can test whether a neutral spatial model is sufficient to reproduce the observed site frequency spectrum. Similarly, quantification of clone density as a function of time~\cite{1_12} and spatial location~\cite{1_14,cancer_sampling1,cancer_sampling2} and number of segregating sites~\cite{Noble2022} can provide orthogonal measurements to reveal whether selection or cell mobility is at play. Recent work has shown that these quantities can be more informative then the clone size distribution to identify selection in boundary-growing tumors~\cite{Noble2022}, and our work provides analytical predictions for the neutral expectation, which can be compared with the experimental data.

\section{Acknowledgements}
We thank Oskar Hallatschek and Ben Simons for helpful discussions. The simulations in this work were performed using resources provided by the Cambridge Service for Data Driven Discovery operated by the University of Cambridge Research Computing Service, provided by Dell EMC and Intel using Tier-2 funding from the Engineering and Physical Sciences Research Council (Capital Grant No. EP/P020259/1), and DiRAC funding from the Science and Technology Facilities Council.

\appendix

\section{Computational Methods} \label{sec:sims}
\begin{figure}[h]
    \centering
    \includegraphics[width=0.45\textwidth]{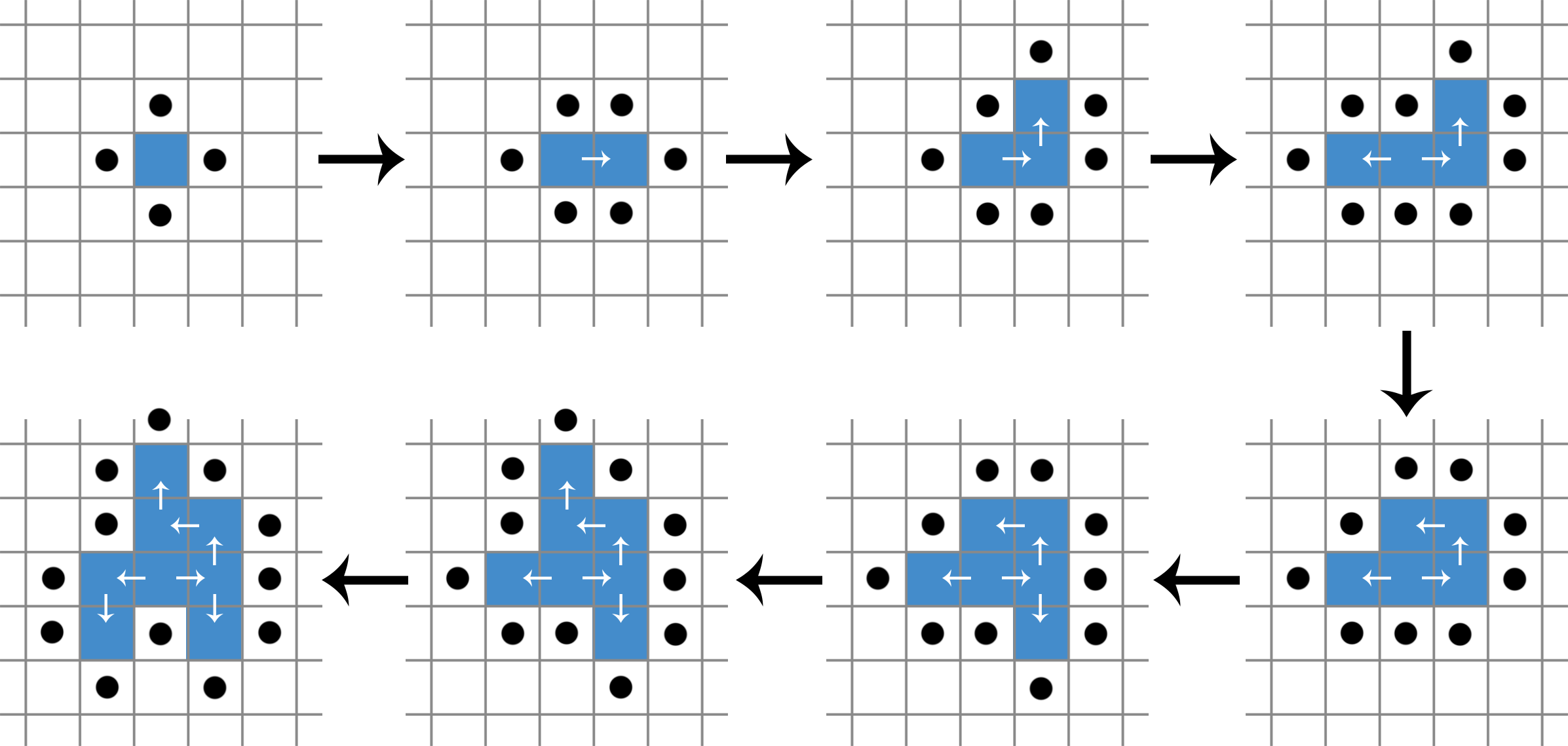}
    \caption{Sketch of how a grid sequentially gets populated via the Eden model. The black dots indicate all possible spots where the new individual could be placed in the next step. The white arrows always point from parent to child.}
    \label{fig:eden_process}
\end{figure}

\begin{figure}[h]
    \centering
    \includegraphics[width=0.45\textwidth]{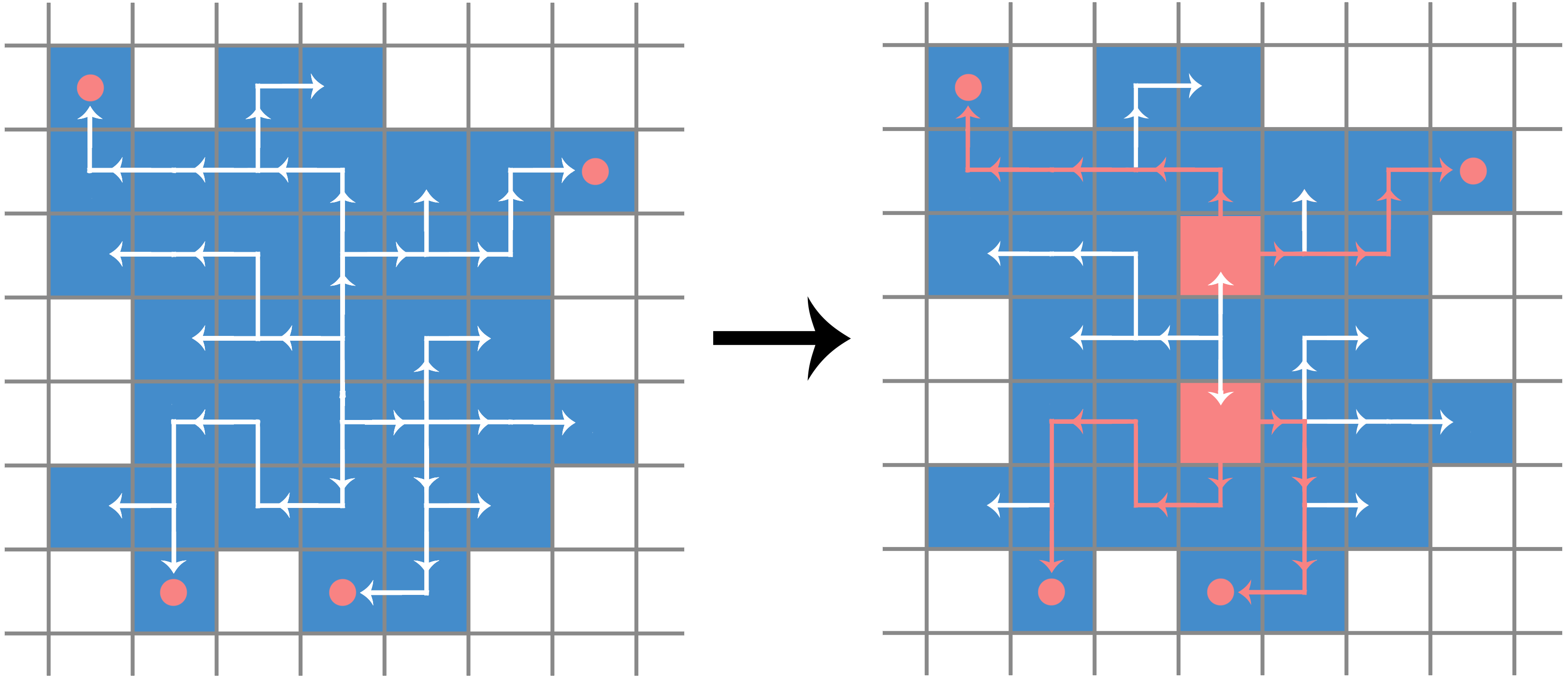}
    \caption{Sketch of how lineages can be traced back to the MRCA. The lineages are traced back for two different cell pairs, one at the top of the colony and one at the bottom, indicated by the red dots. The lineages of the two cells in each pair are traced back, indicated by the red colouring of the lineage lines, until they intersect at the MRCA. The two MRCAs are indicated by the large red squares.}
    \label{fig:mrca_example}
\end{figure}

The computational data was obtained via simulations based on the Eden model \cite{eden}. The growth process in our simulations worked as follows: starting from a beginning set of cells, placed at fixed points on a lattice, we randomly select one cell, which still has at least one free, neighbouring lattice site. We then randomly choose one of these free neighbouring sites and place a new cell there. This new cell can be seen as the descendant of the initially chosen cell. We then repeat the process of randomly choosing a cell and placing a new cell at a neighbouring site, until we reach the desired colony size. An example of how an Eden growth process might look like, can be seen in figure \ref{fig:eden_process}.
By tracking the information about each cell's parent cell, we are able to trace back each cell's lineage and obtain the location of the MRCA of two specific cells. The process for this is shown in figure \ref{fig:mrca_example}.

\begin{figure}[h]
    \centering
    \includegraphics[width=0.45\textwidth]{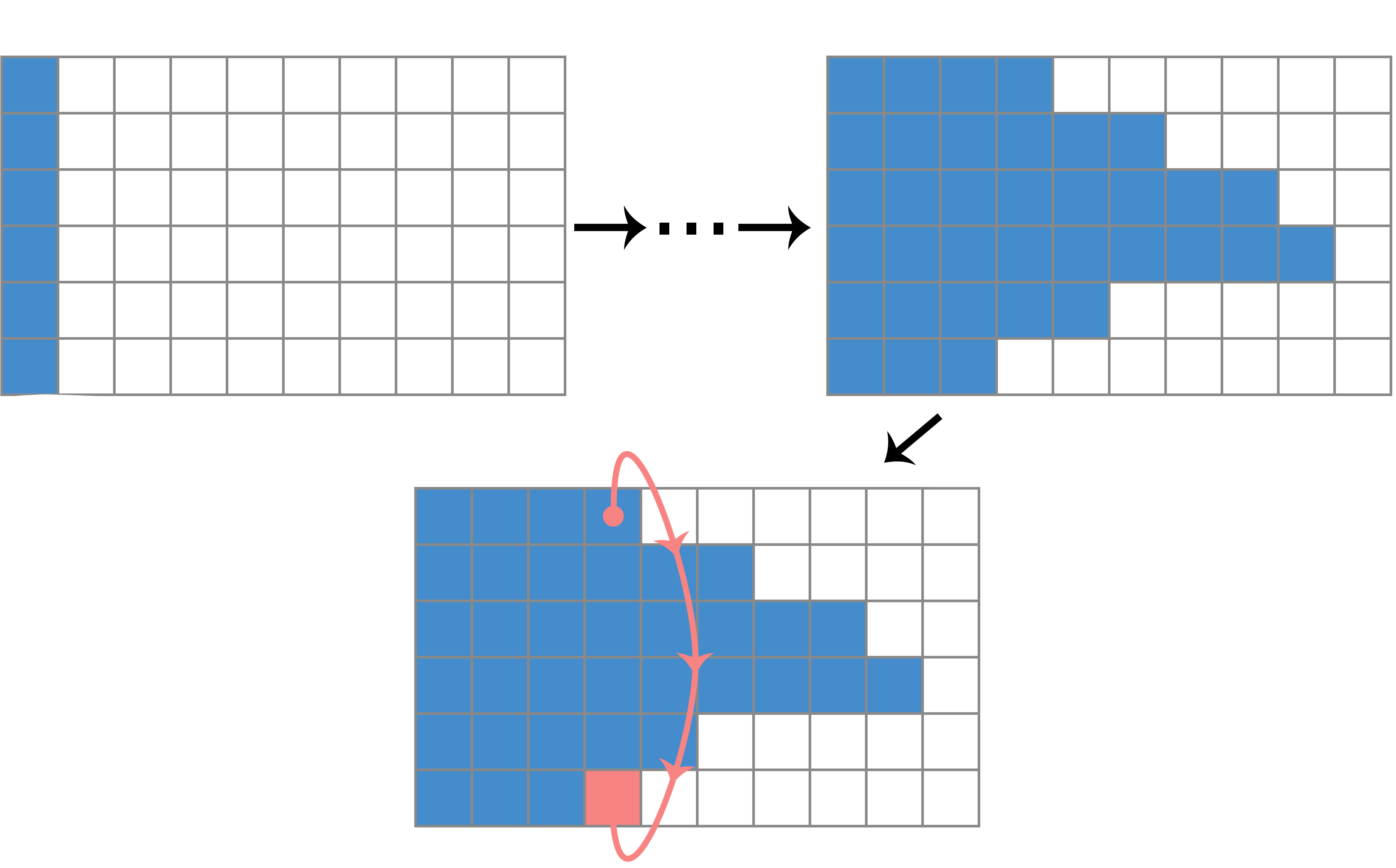}
    \caption{Sketch of an Eden model growth process in a corridor. From the upper left to the upper right picture the corridor gets populated, initially starting from a straight line of individuals filling the whole width of the corridor. The lower picture indicates how a cell at one edge of the corridor can have a descendant at the opposite side, due to the periodic boundary conditions of the corridor.}
    \label{fig:corridor_explanation}
\end{figure}

Figure \ref{fig:eden_process} shows how radial growth was initialized, where we started our simulation from a single cell. In contrast to this, the periodic corridor simulations start with a certain number of cells lying next to each other in a straight line. We then restrict the space that the cells can inhabit to a corridor of a width equal to the number of cells we start with, after which we start the Eden growth process as described above. The corridor boundaries are implemented to be periodical. In other words, the corridor could also be thought of as the outside surface of a cylinder, wrapping around and connecting to itself again. Therefore, a cell which is located at the edge of the corridor can give birth to a new cell at a free spot on the exact opposite side of the corridor. A visualization of this, as well as the growth process in a corridor in general, can be seen in figure \ref{fig:corridor_explanation}.

%In order to keep computation times small, we can used the following optimization method. Throughout the growth process, we keep a list of all cells that still have free neighbouring sites left, i.e. cells which are not completely surrounded by other cells yet. By doing this, when choosing the cell which will reproduce in the next step, we only have to choose from this small list, instead of the whole population. Since the number of cells in this list will be significantly smaller than the number of cells in the entire population, this will lead to much more time efficient computations.

\begin{figure}[h]
    \centering
    \includegraphics[width=0.45\textwidth]{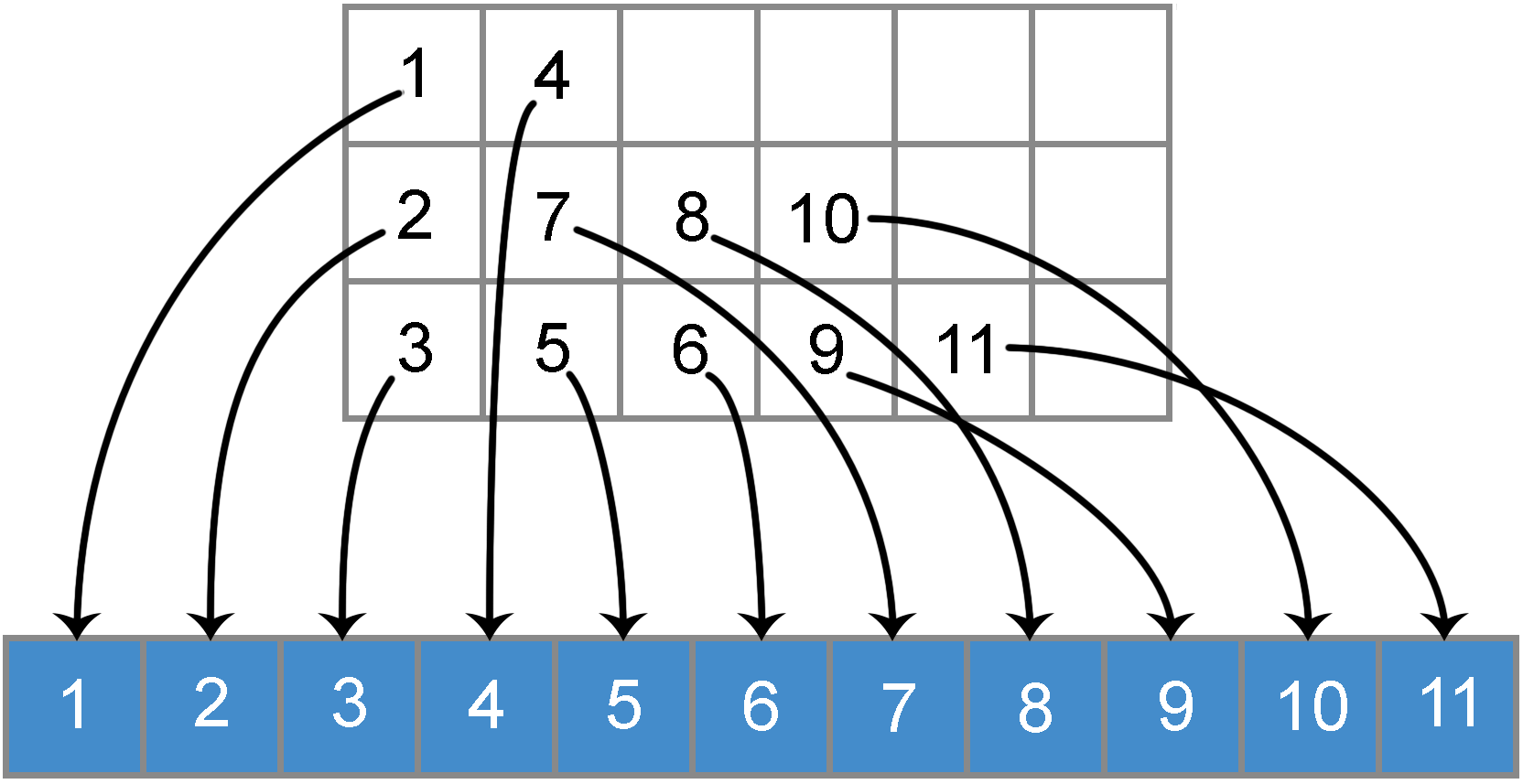}
    \caption{Sketch of the way that information is handled when growing large corridors. The square grid carries pointers to the actual data, which is kept in a data structure (marked in blue) without any spatial relationship between the individual data cells. The grid keeps track of the information about the spatial relationship between the individuals, while the other data structure carries the actual information of the individuals, like the information about their ancestor and children.}
    \label{fig:grid_pointer}
\end{figure}

In order to successfully sample MRCA data from random pairs of cells at the end of the corridor, all of these cells should be descendant of one single ancestor. In contrast to the radial colony, this is not automatically the cause though, since we start our corridor simulation with a column of unrelated cells (whereas in the radial case we start with only a single cell). Due to this, we have to let the colony grow for a significant amount of time until one lineage has pushed out all the others and the cells at the end of the colony are all descendant of one single cell from the very beginning of the corridor. The grid size required to handle corridors with sufficient length for this to happen, greatly exceed the computational memory limitations, even for small corridor widths. Due to this, we implemented a special method for simulating the corridor growth: In order to keep memory requirements low, we should only keep cells in memory which are part of the lineage of the cells at the very front, while deleting all other cells from the memory. At the same time we need to keep track of the spatial structure between the cells at the front, which is necessary for letting new cells grow using the Eden algorithm. To achieve this, we start with a small grid of pointers of the same width as our desired corridor, but a fixed length. Each of the starting cells in our first column has a distinct marker, which all of its descendants will also carry. When growing the grid, we do not save the cells' information on the grid, but rather on an arbitrary data structure (e.g. a list) without any special spatial structure to it. The spatial structure for growing our colony is obtained through the small grid which points to the places in the memory where our cells' information is kept (see figure \ref{fig:grid_pointer}). 
\begin{figure}[h]
    \centering
    \includegraphics[width=0.45\textwidth]{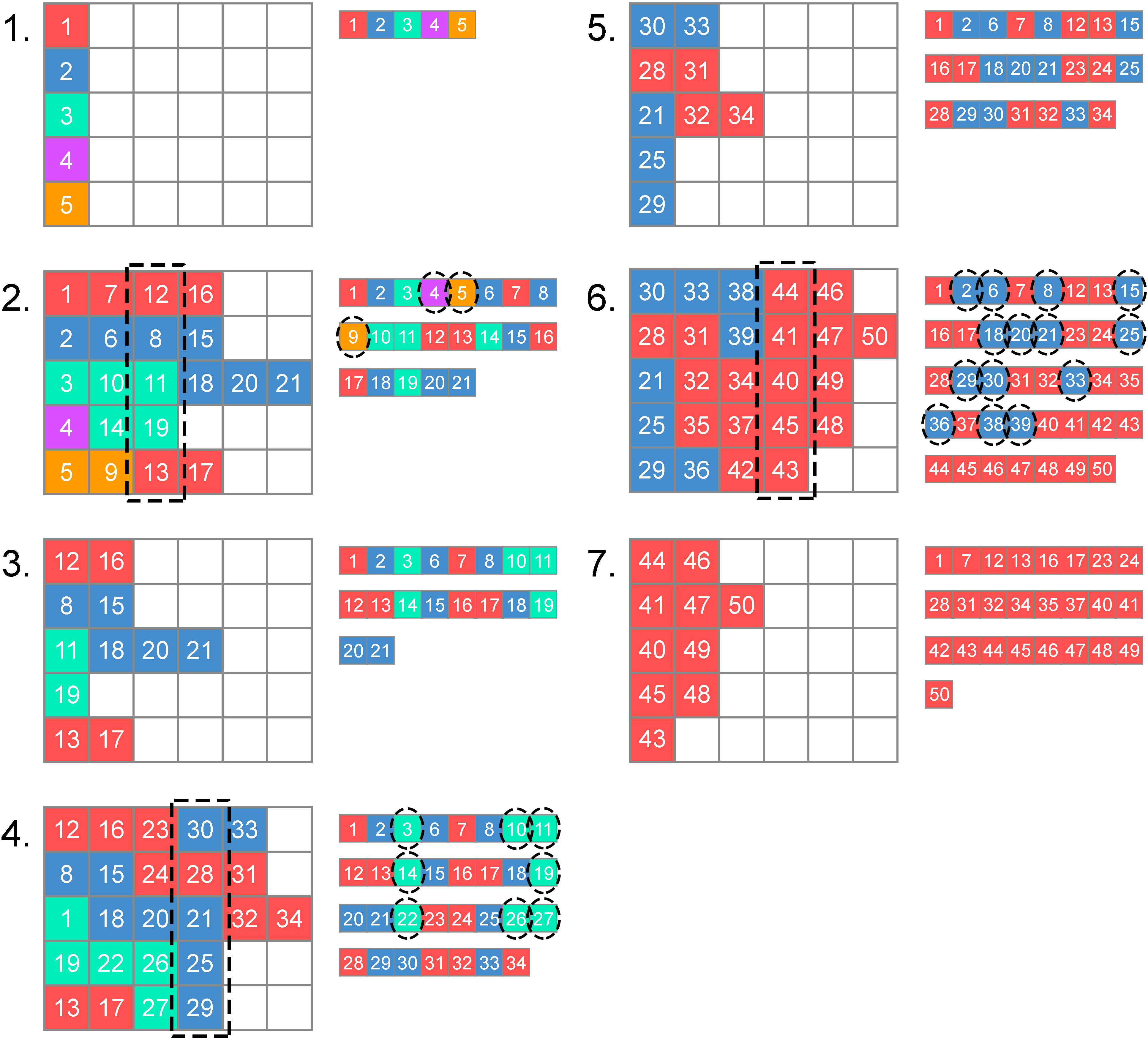}
    \caption{Sketch of the growth process for large corridors, with the spatial data structure carrying the pointers on the left and the data structure carrying the actual cell information on the right.}
    \label{fig:big_cor_process}
\end{figure}
In figure \ref{fig:big_cor_process} we sketched each step of the process for a better understanding. We start with a full column of cells, each with its own marker (see FIG.\ref{fig:big_cor_process}.1). The colony then grows until the first cells hits the right wall, at which point we stop the growth process. We then locate the most right column which is completely filled (marked with a black frame in FIG.\ref{fig:big_cor_process}.2) and check what markers are still present in that column. All cells with markers not present anymore can then be deleted from the data structure on the right (these cells have been circled in FIG.\ref{fig:big_cor_process}.2). After the unnecessary cells have been deleted, we can shift the corridor back to only include columns including and to the right of the column we just investigated for the remaining markers (see FIG.\ref{fig:big_cor_process}.3). From here, the process is repeated again, growing, sampling markers, deleting, until the sampled column only has markers of one kind. At this point we can delete the remaining other cells from our data structure. The colony has now successfully grown to the point that its front only includes cells which are descendant of one single common ancestor (see FIG.\ref{fig:big_cor_process}.4-7). At this point, we can stop growing our colony and start with analyzing its MRCA information.

\section{Lineage density from the universal tree model} \label{sec:cutoff}
Our considerations of the stochastic coalescences with statistics dictated by the KPZ wandering exponent (section~\ref{sec:eden-linear}, together with the incorporation of the inflation effect in the universal tree model (section~\ref{sec:section2}), motivated the following scaling form for the lineage density of radial Eden clusters:
\begin{equation}
  \label{eq:scalingform}
  \frac{l(r)}{2\pi r} = C [x(1-x)]^{-2/3},
\end{equation}
where $x = r/R$ and $C$ is an as-yet undetermined prefactor. The power-law divergences as $x \to 0$ and $x \to 1$  in \eqnref{eq:scalingform} are imposed by the KPZ scaling, but other considerations are needed to fix the value of $C$.

The value of the prefactor $C$ is determined by noting that the above scaling form is not valid out to $x = 1$, but only up to some cutoff distance away from the outer limit of the colony whose value is set by the microscopic details of the growth process. Although the branching process of the universal tree model can be carried out to infinitely many steps, generating ever-finer leaves which approach the outer limit, the true coalescence process is limited by two microscopic length scales in any real biological system or realistic simulation thereof. First, genetic differences do not persist down to infinite  resolution but instead are restricted to some finite spacing. In a microbial colony, for example, the smallest possible spacing between distinct genetic samples is the size of an individual cell; the lineages that are sampled are typically spaced even farther apart. Second, the KPZ wandering statistics arises within a coarse-grained description of the interface between the colony and its environment, which is only valid for roughness features that are larger than some microscopic length scale. This roughness scale is also of the order of a few cells for a microbial expansion.

To fix the value of the prefactor $C$, we impose these length cutoffs at the outer boundary of the colony. We denote the smallest spacing between distinct lineages by the variable $\delta$, and the smallest scale of roughness features by the quantity $\epsilon$. Therefore, the number of distinct samples at the outer boundary is $2\pi R/\delta$. Our proposed scaling form, \eqnref{eq:scalingform}, is only valid out to distances within $\epsilon$ of the outer boundary and we do not expect any more mergers of lineages to occur between $r=R-\epsilon$ and $r=R$. To match the proposed master curve to the number of lineages at the outer boundary, we need
\begin{equation}
  l(R-\epsilon) = C\left[\left(1-\frac{\epsilon}{R}\right) \frac{\epsilon}{R}\right]^{-2/3} \times 2 \pi (R-\epsilon) = \frac{2 \pi R}{\delta}.
\end{equation}
Our scaling arguments are only valid provided $\epsilon/R \ll 1$. Keeping only terms to leading order in $\epsilon/R$ in the above equation, we find
\begin{align}
  2 \pi R C \left(\frac{\epsilon}{R}\right)^{-2/3} &= \frac{2 \pi R}{\delta} \\
  \Rightarrow C = \frac{\epsilon^{2/3}}{R^{2/3}\delta}.\label{eq:C}
\end{align}

Equation~\eqref{eq:C} specifies the prefactor by requiring the proposed master curve to match the lineage density at the edge of the colony. In the Eden simulations used in our work, both length scales $\epsilon$ and $\delta$ are given by the lattice spacing, hence we can set $\delta = \epsilon$ which gives
$$C = (R^2\epsilon)^{-1/3}$$
up to some $O(1)$ constant which we assume to be one. This value of the prefactor gives rise to the complete expression in \eqnref{eq:density_th}. We find that the expression gives a reasonable match to the rescaled lineage density measurements from Eden simulations (compare dashed line to solid lines in \figref{fig:collapsed}) without any fits being performed. The agreement could be slightly improved by treating the $O(1)$ numerical constant as a free parameter whose value is determined by fitting the proposed master curve to the data. If we were to analyze genealogical tree data from a biological population, we would not have microscopic information about the quantities $\epsilon$ and $\delta$. In that case, it would be appropriate to use \eqnref{eq:C} and fix the combination $\epsilon^{2/3}/\delta$ as a fitting parameter.

\bibliography{apssamp}% Produces the bibliography via BibTeX.

\end{document}